\documentclass[12pt]{article}
\usepackage{eurosym}
\usepackage{graphicx}
\usepackage[utf8]{inputenc}
\usepackage[T1]{fontenc}
\usepackage{indentfirst}
\usepackage[margin=0.6in,nomarginpar]{geometry}
\usepackage[final]{hyperref}
\usepackage{amsmath}
\usepackage{hyperref}
\usepackage{cite}
\usepackage{subcaption}
\usepackage{caption}
\usepackage{amssymb}
\usepackage{multirow}
\usepackage[table,dvipsnames]{xcolor}
\usepackage{orcidlink}
\usepackage{float}

\hypersetup{
colorlinks=true,
linkcolor=blue,
citecolor=blue,
filecolor=magenta,
urlcolor=blue
}
\begin{document}

\title{Uncertainty Relation for Pseudo-Hermitian Quantum Systems}


\author{Boubakeur Khantoul\thanks{ boubakeur.khantoul@univ-constantine3.dz}
\\Department of Process Engineering, University of \\ Constantine 3 - Salah Boubnider, postcode 25016, Constantine, Algeria. \\
Theoretical Physics Laboratory, Department of Physics, University of Jijel, Algeria.\and Bilel Hamil\thanks{hamilbilel@gmail.com}
\\Laboratoire de Physique Mathématique et Subatomique, Faculté des Sciences \\ Exactes, Université Constantine 1,Constantine, Algeria. \and Amar Benchikha\thanks{benchikha4@yahoo.fr}
\\Département de TC, Faculté de SNV, Université Constantine 1, \\Constantine, Algeria.
}
\maketitle
\begin{abstract}
This study investigates pseudo-Hermitian quantum mechanics, where the Hamiltonian satisfies a modified Hermiticity condition. We extend the uncertainty relation for such systems, demonstrating its equivalence to the standard Hermitian case within a pseudo-Hermitian inner product. Analytical solutions to the time-dependent Schrödinger equation with a linearly evolving potential are derived. Furthermore, we show that the uncertainty relation for position and momentum remains real and greater than $\frac{1}{2}$, highlighting the significance of non-Hermitian systems in quantum mechanics.
\end{abstract}



\maketitle

\section{Introduction}\label{sec1}

Non-Hermitian quantum mechanics emerged from the discovery that certain
non-Hermitian Hamiltonians can possess entirely real spectra under specific
conditions, particularly when they exhibit $\mathcal{PT}$-symmetry.
Introduced by Carl Bender and Stefan Boettcher in the late 1990s, $\mathcal{%
PT}$-symmetric quantum mechanics posits that a Hamiltonian $H$ can have real
eigenvalues if it is invariant under combined parity ($\mathcal{P}$) and
time-reversal ($\mathcal{T}$) transformations, even though $H$ itself is not
Hermitian in the traditional sense \cite{2,3,4,5}. This groundbreaking idea broadened the conventional framework of quantum mechanics, allowing for
a broader class of Hamiltonians and offering new perspectives on quantum
theory.

The concept of pseudo-Hermiticity plays a pivotal role in this context,
providing a broader framework that encompasses $\mathcal{PT}$-symmetry as a
special case. A Hamiltonian is termed pseudo-Hermitian if it satisfies a specific condition involving its Hermitian conjugate, mediated by a linear, invertible, and Hermitian metric operator $\eta $%
\begin{equation}
H^{\dagger }=\eta H\eta ^{-1}.  \label{1a}
\end{equation}%
This relationship ensures that, even though the Hamiltonian itself is not
Hermitian, it possesses a real spectrum under certain conditions, bridging
the gap between non-Hermitian and conventional quantum mechanics. The associated scalar product remains conserved over time and is defined via the metric operator as $\left\langle .\right\vert .\rangle _{\eta
}=\left\langle .\right\vert \eta \left\vert .\right\rangle .$ Additionally,
every pseudo-Hermitian Hamiltonian can be represented as a diagonalizable
operator with a spectrum consisting of real numbers. Consequently, it can be
transformed into a Hermitian operator, denoted $h$, through a similarity
transformation expressed as $h=\rho H\rho ^{-1}$, where the operator $\rho $
is linear and invertible and is defined by $\eta =\rho \rho ^{\dag }$.
Notably, the Hamiltonians $h$ and $H$ possess the same real eigenvalue spectra , and their corresponding eigenfunctions $\left\vert \psi
\right\rangle $\ and $\left\vert \Psi \right\rangle $ respectively, are
linked by the relationship $\left\vert \psi \right\rangle =\rho \left\vert
\Psi \right\rangle $ \cite{6,7,8}.
The implications of non-Hermitian quantum systems are profound and
wide-ranging. In optics and photonics, symmetric structures $\mathcal{PT}$
have led to the development of innovative devices that exploit balanced gain
and loss to achieve unique optical properties, such as unidirectional
invisibility and loss-induced transparency \cite{8m}. These phenomena have
practical applications in optical waveguides, lasers, and sensors.
In condensed matter physics, non-Hermitian Hamiltonians have been
instrumental in exploring new topological phases of matter, providing a
richer understanding of topological properties and potential applications in
fault-tolerant quantum computing and robust electronic devices. Furthermore,
in quantum field theory and particle physics, non-Hermitian models describe
systems where particles interact with an environment, leading to decay
processes and open quantum systems, which are crucial for understanding
phenomena such as resonances and the behavior of quasiparticles in
high-energy physics and condensed matter systems \cite{9m,10m,11m,12m}.

Despite their successes, non-Hermitian quantum systems are not without
significant challenges. One of the primary issues is related to the
uncertainty relation in the context of non-Hermitian Hamiltonians. In
standard quantum mechanics, the uncertainty principle is a fundamental
concept that imposes limits on the precision with which pairs of conjugate
variables, such as position and momentum, can be simultaneously known. This
principle is inherently linked to the Hermiticity of the operators
representing these variables, ensuring real and physically meaningful
uncertainties.

In non-Hermitian quantum systems, however, the situation becomes more
complex. When a non-Hermitian Hamiltonian is considered, the inner product
and the associated Hilbert space structure must be redefined to ensure a
consistent probabilistic interpretation. The standard inner product in Hilbert space may not be suitable, requiring an alternative inner product that preserves pseudo-Hermiticity or $\mathcal{PT}$%
-symmetry of the Hamiltonian. This redefinition poses a significant problem
for the uncertainty relation \cite%
{25,26,27}. In the new inner product space, operators that
were Hermitian in the original Hilbert space are generally no longer
Hermitian. Thus, applying the standard uncertainty principle may yield complex values, which lack physical significance since uncertainty must correspond to a real, measurable quantity describing fundamental measurement limitations.

To resolve this issue, we formulate an uncertainty relation adapted to the pseudo-Hermitian framework, extending the conventional Hermitian approach while establishing a clear correspondence between them. These formulations involve redefining the commutation relations and expectation
values within a new inner product space, ensuring that the derived
uncertainties are real and physically meaningful. Furthermore, we clearly demonstrate that the uncertainty relation in the new Hilbert space, characterized by the metric operator $\eta$, is equivalent to the standard uncertainty relation in the conventional Hilbert space.

As an application, we solved the time-dependent Schr\"{o}dinger equation (TDSE) for a particle confined within a time-dependent complex triangular potential well using the Lewis-Riesenfeld invariant method. This method involves constructing an invariant operator that commutes with the Hamiltonian at all times, enabling us to derive exact solutions for the system's wave function.
Using this technique, we derived the eigenstates and their time-dependent phases, enabling a detailed investigation of quantum dynamics in a complex, time-dependent potential.

Subsequently, we examined the pseudo-Hermitian uncertainty relation for this
system. Our findings revealed that the value of the uncertainty relation was
real and greater than $1/2$, thus verifying the consistency of our formulations. This result demonstrates that the Heisenberg uncertainty
principle holds true within the pseudo-Hermitian framework, reinforcing the
physical validity of our approach and highlighting the robustness of the
uncertainty relations in these generalized quantum systems.

\section{Uncertainty Relation in Pseudo-Hermitian Inner Product}\label{sec2}

The uncertainty relation within the framework of pseudo-Hermitian inner
product significantly modifies the standard quantum mechanics approach. In the conventional setting, where the observables are described by
Hermitian operators, Heisenberg Uncertainty Principle governs the relationship between uncertainties in measurements of incompatible observables. For observables represented by Hermitian operators $A$ and $B$, the principle is articulated as
\begin{equation}
\left( \triangle A\right) ^{2}.\left( \triangle B\right) ^{2}\geq\frac{1}{4}\left\vert \left\langle
\varphi\right\vert \left[ A,B\right] \left\vert \varphi\right\rangle
\right\vert ^{2},  \label{a1}
\end{equation}
where $\triangle A$ and $\triangle B$ represent the uncertainty in
measurements of observable $A$ and $B$, given by 
\begin{equation}
\left( \triangle A\right) ^{2}=\left\langle \phi\right\vert A^{2}\left\vert \phi
\right\rangle -\left\langle \phi\right\vert A\left\vert \phi\right\rangle
^{2},  \label{a2}
\end{equation}%
where $\left\vert \phi\right\rangle $ is a state in the usual Hilbert space,
with A replaced by B for $\triangle B $.

However, in systems governed by pseudo-Hermitian Hamiltonians, the
introduction of a metric operator $\eta$ requires a modification of the inner product used for calculating expectation values and uncertainties, ensuring the physical consistency of the theory. Consequently, the uncertainty relation within this framework accounts for this adjusted inner product.

In the context of pseudo-Hermitian systems, the operators
undergo transformations. Specifically, the operators $A$ and $B$ are
replaced by pseudo-Hermitian operators $\widetilde{A}=\rho ^{-1}A\rho $ and $%
\widetilde{B}=\rho ^{-1}B\rho $\bigskip , respectively.

Additionally, the state $\left\vert \phi \right\rangle $ undergoes transformation as $%
\left\vert \phi \right\rangle =\rho \left\vert \psi \right\rangle $.
Following this transformation, it becomes apparent that 
\begin{equation}
\left\langle \phi \right\vert A^{2}\left\vert \phi \right\rangle
=\left\langle \psi \right\vert \rho ^{\dagger }AA\rho \left\vert \psi
\right\rangle =\left\langle \psi \right\vert \rho ^{\dagger }AA\rho \left\vert
\psi \right\rangle =\left\langle \psi \right\vert \left( \widetilde{A}%
\right) ^{\dagger }\eta \widetilde{A}\left\vert \psi \right\rangle ,  \label{a4}
\end{equation}%
and 
\begin{equation}
\left\langle \phi \right\vert A\left\vert \phi \right\rangle
^{2}=\left\langle \psi \right\vert \rho ^{\dag }A\rho \left\vert \psi
\right\rangle ^{2}=\left\langle \psi \right\vert \eta \widetilde{A}%
\left\vert \psi \right\rangle ^{2},  \label{a5}
\end{equation}%
where $\eta =\rho ^{\dag }\rho .$ Since $\widetilde{A}$ is quasi-Hermitian, $%
\eta \widetilde{A}=\left( \widetilde{A}\right) ^{\dag }\eta ,$ then 
\begin{equation}
\left\langle \phi \right\vert A\left\vert \phi \right\rangle
^{2}=\left\langle \psi \right\vert \eta \widetilde{A}\left\vert \psi
\right\rangle ^{2}=\left\langle \psi \right\vert \eta \widetilde{A}%
\left\vert \psi \right\rangle \left\langle \psi \right\vert \left( 
\widetilde{A}\right) ^{\dag }\eta \left\vert \psi \right\rangle.   \label{a6}
\end{equation}%

The standard deviation of $A$ (with $A$ replaced by $B$ in $\left( \triangle B\right) ^{2}$)
is transformed as follows %
\begin{equation}
\left( \triangle A\right) ^{2}=\left( \triangle \widetilde{A}_{\eta }\right)
^{2}=\left\langle \psi \right\vert \left( \widetilde{A}\right) ^{\dagger }\eta 
\widetilde{A}\left\vert \psi \right\rangle -\left\langle \psi \right\vert
\eta \widetilde{A}\left\vert \psi \right\rangle \left\langle \psi
\right\vert \left( \widetilde{A}\right) ^{\dagger }\eta \left\vert \psi
\right\rangle .  \label{a7}
\end{equation}%

Consequently, the modified Heisenberg Uncertainty Principle for two
non-commuting observables $\tilde{A}$ and $\tilde{B}$ within this altered
inner product framework is expressed as%
\begin{equation}
\left( \triangle \widetilde{A}_{\eta }\right) ^{2}\left( \triangle 
\widetilde{B}_{\eta }\right) ^{2}=\left( \triangle A\right) ^{2}\left( \triangle B\right) ^{2}\geq 
\frac{1}{4}\left\vert \left\langle \psi \right\vert \eta \left[ \widetilde{A}%
,\widetilde{B}\right] \left\vert \psi \right\rangle \right\vert ^{2}.
\label{a9}
\end{equation}%

In this section, we have explored how the uncertainty relation is modified
within the pseudo-Hermitian inner product framework. By introducing a metric
operator $\eta $, we adapted the Heisenberg Uncertainty Principle to account
for pseudo-Hermitian operators and states. This reformulation ensures that
the uncertainties derived are physically meaningful and consistent with the
modified inner product space. The resulting uncertainty relation
demonstrates a significant extension of the conventional quantum mechanics
approach, providing deeper insights into the behavior of non-Hermitian
systems.

\section{A particle confined within a TD complex triangular potential well}\label{sec3}

The introduction of time-dependent non-Hermitian Hamiltonians requires new approaches to solve the TDSE, among which the invariant method emerges
as a powerful tool \cite{13a,9, 10, 11, 12, 13, 14, 16, 17, 18, 19, 20, 21,
22, 23, 24}. By constructing time-dependent invariants that adhere to the
pseudo-Hermitian structure, one can formulate solutions to the TDSE that
accurately capture the dynamics of non-Hermitian systems, offering insights
into phenomena such as quantum state evolution and transition probabilities
in these unconventional settings.

We consider the Hamiltonian $H\left( t\right) $ of a particle with a time-dependent mass confined within a complex time-dependent triangular potential well, which is given by%
\begin{equation}
V\left( x,t\right) =\left\{ 
\begin{array}{c}
if(t)x,\text{ \ if \ \ \ \ \ \ }x>0 \\ 
\infty,\text{\ \ \ \ \ \ \ if \ \ \ \ \ \ }x\leq0%
\end{array}
\right. .  \label{99}
\end{equation}

This model is particularly useful as it provides a simpler mathematical form while still capturing essential physics in scenarios where the
potential varies in a non-stationary manner. The TDSE that describes this system for $x>0$ is given by%
\begin{equation}
\frac{p^{2}}{2m\left( t\right) }\Psi\left( x,t\right) +if(t)x\Psi\left(
x,t\right) =i\frac{\partial}{\partial t}\Psi\left( x,t\right) ,  \label{100}
\end{equation}
where $m\left( t\right) $ and $f(t)$ are arbitrary real time dependent
functions, we note that for all the rest of the paper we take $\hbar=1$.

According to \cite{13}, if a non-trivial pseudo-Hermitian invariant $I_{ph}$
exists and satisfies the Von Neumann equation \cite{13a}, a solution of the
Schr\"{o}dinger equation with a time-dependent complex Hamiltonian is easily
found.

\begin{equation}
\frac{dI_{ph}(t)}{dt}=\frac{\partial I_{ph}(t)}{\partial t}-i%
\left[ I_{ph}\left( t\right) ,H\left( t\right) \right] =0.  \label{VN}
\end{equation}%

To solve TDSE $\left( \ref{100}\right) ,$ we search for a time-dependent
invariant of the following form%
\begin{equation}
I_{ph}\left( t\right) =\gamma _{1}\left( t\right) p^{2}+\gamma _{2}\left(
t\right) x+\gamma _{3}\left( t\right) p+\gamma _{4}\left( t\right) ,
\label{101}
\end{equation}%
where $\gamma _{i}\left( t\right) $ are arbitrary complex functions to be defined.

By substituting expressions $%
\left( \ref{100}\right) $ and $\left( \ref{101}\right) $ into $\left( \ref%
{VN}\right),  $ 
we obtain the following system of equations

\begin{align}
\dot{\gamma}_{1}\left( t\right) & =0,  \notag \\
\dot{\gamma}_{2}\left( t\right) & =0,  \label{103} \\
\dot{\gamma}_{3}\left( t\right) & =-\frac{\gamma_{2}(t)}{m\left( t\right) }%
+2if\left( t\right) \gamma_{1}(t),  \notag \\
\dot{\gamma}_{4}\left( t\right) & =if\left( t\right) \gamma _{3}(t).  \notag
\end{align}

Since the invariant is not unique, we choose a specific form to simplify the calculations, where we take $\gamma_{1}(t)=\gamma_{2}(t)=1,$ then $\gamma_{3}(t)$
and $\gamma _{4}(t)$ are given by%
\begin{equation}
\gamma_{3}(t) =ia( t) -b( t) \text{ \ \ \ where \ \ }%
a( t) =2\int f( t) dt\text{ \ \ and }b( t)
=\int\frac {dt}{m(t) },  \label{104}     
\end{equation}
\begin{equation}
\gamma_{4}(t)  =-c\left( t\right) -id\left( t\right) \text{ \ where }%
c\left( t\right) =2\int f\left( t\right) a\left( t\right) dt\text{ \ and }%
d\left( t\right) =\int f\left( t\right) b\left( t\right) dt.  \label{105}
\end{equation}

Substituting the Eqs. $\left( \ref{104}\right) $ and $\left( \ref{105}%
\right) $ in the Eq. $\left( \ref{101}\right) $ we found%
\begin{equation}
I_{ph}\left( t\right) =p^{2}+x+\left[ ia\left( t\right) -b\left( t\right) %
\right] \text{ }p-c\left( t\right) -id\left( t\right),  \label{106}
\end{equation}
when its eigenvalue equation is presented as%
\begin{equation}
I_{ph}\left( t\right) \left\vert \psi\left( x,t\right) \right\rangle
=\lambda_{n}\left\vert \psi\left( x,t\right) \right\rangle .  \label{107}
\end{equation}

To show the reality of the eigenvalues of $I_{ph}\left( t\right) $, we seek a metric operator that satisfies the pseudo-Hermiticity relation%
\begin{equation}
I_{ph}^{\dagger}\left( t\right) =\eta\left( t\right) I_{ph}\left( t\right)
\eta^{-1}\left( t\right) .  \label{108}
\end{equation}

To solve the equation $\left( \ref{108}\right) $ and given that the metric operator is not unique, we make the following selection%
\begin{equation}
\eta\left( t\right) =\exp[\alpha\left( t\right) x+\beta\left( t\right) p],
\label{109}
\end{equation}
Where $\alpha(t)$ and $\beta(t)$ are real time-dependent functions, these functions can be explicitly determined using the Baker-Campbell-Hausdorff (BCH) formula as follows:
\begin{align}
\alpha\left( t\right) & = -a\left( t\right), \label{111} \\
\beta\left( t\right) & = a\left( t\right) b\left( t\right) - 2d\left( t\right). \label{112}
\end{align}
then the TD metric $\eta\left( t\right) $ is given by%
\begin{equation}
\eta\left( t\right) =\exp[-a\left( t\right) x+(a\left( t\right) b\left(
t\right) -2d\left( t\right) )p].  \label{113}
\end{equation}
For simplicity, we assume that $\eta\left( t\right) = \rho^{2}\left( t\right)$, where $\rho\left( t\right)$ is a Hermitian operator. The expression for $\rho\left( t\right)$ is given by: 
\begin{equation}
\rho\left( t\right) = \exp\left[ \frac{-a\left( t\right)}{2}x + \left(\frac{a\left( t\right) b\left( t\right)}{2} - d\left( t\right)\right)p\right],  
\label{114}
\end{equation}
and the Hermitian invariant $I_{h}$, associated with the pseudo-Hermitian invariant $I_{ph}$, is given by:  
\begin{equation}
I_{h}\left( t\right) = \rho\left( t\right) I_{ph}\left( t\right) \rho^{-1}\left( t\right) 
= p^{2} + x - b\left( t\right)p + \frac{a^{2}\left( t\right)}{4} - c\left( t\right).  
\label{115}
\end{equation}
The eigenvalue equation of $I_{h}\left( t\right) $ is given by%
\begin{equation}
I_{h}\left( t\right) \left\vert \phi\left( x,t\right) \right\rangle
=\lambda_{n}\left\vert \phi\left( x,t\right) \right\rangle .  \label{116}
\end{equation}
Since $I_{h}\left( t\right)$ and $I_{ph}\left( t\right)$ are quasi-Hermitian operators, they are therefore iso-spectral. Moreover, the eigenfunctions of $I_{ph}\left( t\right)$ and $I_{h}\left( t\right)$ are connected by the following relation:
\begin{equation}
\left\vert \psi\left( x,t\right) \right\rangle =\rho^{-1}\left( t\right)
\left\vert \phi\left( x,t\right) \right\rangle .  \label{117}
\end{equation}

To derive the eigenvalues of the invariant operator in Eq. $(\ref{116})$, we
introduce the unitary transformation $U(t)$ as%
\begin{equation}
\left\vert \phi\left( x,t\right) \right\rangle =U(t)\left\vert \varphi\left(
x,t\right) \right\rangle ,  \label{118'}
\end{equation}
where%
\begin{equation}
U(t)=\exp\left[ i\frac{b\left( t\right) }{2}x+\frac{i}{4}\left[ a^{2}\left(
t\right) -b^{2}\left( t\right) -4c\left( t\right) \right] p\right] .
\label{118}
\end{equation}
According to this transformation and the Baker-Campbell-Hausdorff (BCH) formula, the Hermitian TD invariant $I_{h}\left( t\right)$ is transformed as:  
\begin{equation}
U(t) I_{ph} U^{-1}(t) = I_{h} = p^{2} + x.  
\label{119}
\end{equation}
This transformation simplifies the problem of solving a time-dependent eigenvalue equation into a time-independent equation with a known solution \cite{28,29}.

\begin{equation}
\left\vert \varphi_{n}\left( x,t\right) \right\rangle =\frac{1}{Ai^{\prime
}\left( a_{n}\right) }Ai\left( x+a_{n}\right) ,  \label{120}
\end{equation}
and the eigenvalues $\lambda_{n}$ are reals and discrete determined by the zeros of Airy function \cite{30}%
\begin{equation}
\lambda_{n}=-a_{n+1}.  \label{121}
\end{equation}

Substituting the Eqs. (\ref{118}) and (\ref{120}) in Eq. (\ref{118'}), we
find the solution of the eigenvalues equation of the Hermitian invariant $%
I_{h}(t)$%
\begin{equation}
\left\vert \phi_{n}\left( x,t\right) \right\rangle =\frac{1}{Ai^{\prime
}\left( a_{n}\right) }U\left( t\right) Ai\left( x+a_{n}\right) ,  \label{122}
\end{equation}
in the other hand, by inserting the Eqs. $\left( \ref{122}\right) $ and $%
\left( \ref{114}\right) $\ in $\left( \ref{117}\right) ,$ the solution of
the Eq $\left( \ref{107}\right) $ is written in the form%
\begin{equation}
\left\vert \psi\left( x,t\right) \right\rangle =\frac{1}{Ai^{\prime}\left(
a_{n}\right) }\rho^{-1}\left( t\right) U\left( t\right) Ai\left(
x+a_{n}\right) .  \label{123}
\end{equation}

 Since the solution of the time-dependent Schr\"{o}dinger equation $\left(\ref{100}\right)$ is connected with the eigenstate of the pseudo-Hermitian invariant through a time-dependent phase factor, we can write:
\begin{equation}
\left\vert \Psi \left( x,t\right) \right\rangle = e^{i\epsilon _{n}\left(t\right)}\left\vert \psi \left( x,t\right) \right\rangle, \label{124}
\end{equation}
where the phase $\epsilon_{n}\left( t\right)$ is calculated using the following relation:  
\begin{align}
\dot{\epsilon}_{n}\left( t\right) &= \left\langle \psi(x,t)\right\vert \eta(t) \left[ i\frac{\partial}{\partial t} - H(t) \right] \left\vert \psi(x,t)\right\rangle.  
\label{a125}
\end{align}
Using Eqs. $\left(\ref{114}\right)$, $\left(\ref{118'}\right)$, $\left(\ref{119}\right)$, and $\left(\ref{120}\right)$, we find:
\begin{equation}
\epsilon_{n}\left( t\right) = \int \left( \theta\left( t\right) - \chi\left(t\right) - \frac{\lambda_{n}}{2m\left( t\right) }\right) dt, \label{133}
\end{equation}
where
\begin{equation}
\theta \left( t\right) = -\frac{f\left( t\right)}{2}\left[ \frac{a\left(t\right)b\left( t\right)}{2} - \int f\left( t\right) b\left( t\right) dt \right], \label{128}
\end{equation}
and
\begin{equation}
\chi \left( t\right) = \frac{1}{16m\left( t\right)}\left[ a^{2}\left(t\right) - b^{2}\left( t\right) - 4c\left( t\right) \right]. \label{131'}
\end{equation}
Thus, the solution of the time-dependent Schr\"{o}dinger equation $\left(\ref{100}\right)$ is given by:
\begin{equation}
\Psi\left( x,t\right) = \frac{1}{Ai^{\prime}\left( a_{n}\right)} \exp\left[i\epsilon_{n}\left( t\right) \right] \rho^{-1}\left( t\right) U\left(t\right) Ai\left( x + a_{n} \right). \label{134}
\end{equation}

The action of the transformation $\rho^{-1}\left( t\right) $ on the wave
function $\left\vert \phi\left( x,t\right) \right\rangle $ in $x$%
-representation is given 
\begin{equation}
\left\langle x\left\vert \rho_{j}\left( t\right) ^{-1}\right\vert \phi
_{j}\left( x,t\right) \right\rangle =\exp\left[ i\zeta\left( t\right) \right]
\exp\left[ \frac{a\left( t\right) }{2}x\right] \left\vert \phi\left(
x+i\left( \frac{a\left( t\right) b\left( t\right) }{2}-d\left( t\right)
\right) ,t\right) \right\rangle ,  \label{135}
\end{equation}
where 
\begin{equation}
\zeta\left( t\right) =\frac{a\left( t\right) }{4}\left( \frac{a\left(
t\right) b\left( t\right) }{2}-d\left( t\right) \right) .  \label{136}
\end{equation}
Finally, the solution of the TDSE $\left( \ref{100}%
\right) $ can be written as 
\begin{equation}
\left\vert \Psi\left( x,t\right) \right\rangle =\exp\left[ i\left(
\epsilon_{n}\left( t\right) +\zeta\left( t\right) \right) \right] \exp\left[ 
\frac{a\left( t\right) }{2}x\right] \left\vert \phi\left( x+i\left( \frac{%
a\left( t\right) b\left( t\right) }{2}-d\left( t\right) \right) ,t\right)
\right\rangle .  \label{137}
\end{equation}

In quantum mechanics, determining the expected values of
crucial operators, such as position $x$ and momentum $p$ is essential to understand the behavior of physical systems. These operators $x$ and $p$%
are transformed into $X=\rho^{-1}x\rho$ and $P=\rho^{-1}p\rho$,
respectively. In particular, the transformation ensures that $x$ and $X$ share the
same mean value, as do $p$ and $P$.

To determine the standard deviation of $X$, the expressions are as follows: 
\begin{align}
\left\langle \Psi \right\vert \eta X\left\vert \Psi \right\rangle &
=\left\langle \phi \right\vert x\left\vert \phi \right\rangle =\left\langle
\varphi \right\vert x\left\vert \varphi \right\rangle -\frac{1}{4}\left(
a^{2}-b^{2}-4c\right)   \notag \\
& =\frac{-2}{3}a_{n}-\frac{1}{4}\left( a^{2}-b^{2}-4c\right) ,  \label{138}
\end{align}%
and%
\begin{align}
\left\langle \Psi \right\vert X^{\dag }\eta X\left\vert \Psi \right\rangle &
=\left\langle \phi \right\vert x^{2}\left\vert \phi \right\rangle
=\left\langle \varphi \right\vert x^{2}\left\vert \varphi \right\rangle -%
\frac{1}{2}\left( a^{2}-b^{2}-4c\right) \left\langle \varphi \right\vert
x\left\vert \varphi \right\rangle +\frac{1}{16}\left( a^{2}-b^{2}-4c\right)
^{2}  \notag \\
& =\frac{8}{15}a_{n}^{2}+\frac{1}{3}a_{n}\left( a^{2}-b^{2}-4c\right) +\frac{%
1}{16}\left( a^{2}-b^{2}-4c\right) ^{2}.  \label{139}
\end{align}

Thus, the standard deviation of $X$ is 
\begin{equation}
\left( \triangle X\right) ^{2}=\left( \triangle x\right) ^{2}=\frac{4}{45}a_{n}^{2}.  \label{140}
\end{equation}

For the operator $P$ we have%
\begin{equation}
\left\langle \Psi \right\vert \eta P\left\vert \Psi \right\rangle
=\left\langle \phi \right\vert p\left\vert \phi \right\rangle =\left\langle
\varphi \right\vert p\left\vert \varphi \right\rangle +\frac{b}{2}=\frac{b}{2%
},  \label{140b}
\end{equation}%
where $\left\langle \varphi \right\vert p\left\vert \varphi \right\rangle =0.$ 

On the other hand
\begin{align}
\left\langle \Psi \right\vert P^{\dagger }\eta P\left\vert \Psi \right\rangle &
=\left\langle \phi \right\vert p^{2}\left\vert \phi \right\rangle
=\left\langle \varphi \right\vert p^{2}\left\vert \varphi \right\rangle
+b\left\langle \varphi \right\vert p\left\vert \varphi \right\rangle +\frac{%
b^{2}}{4}  \notag \\
& =-\frac{a_{n}}{3}+\frac{b^{2}}{4},  \label{141}
\end{align}%
consequently, the standard deviation of $P$ is%
\begin{equation}
\left( \triangle P\right) ^{2}=\left( \triangle p\right) ^{2}=-\frac{a_{n}}{3}.  \label{142}
\end{equation}

Finally, the uncertainty relation is expressed as 
\begin{equation}
\triangle X\triangle P=\triangle x\triangle p=\sqrt{-\frac{4}{135}a_{n}^{3}}>%
\frac{1}{2}.  \label{143}
\end{equation}

Because $a_n$ are negative,  this result is both real and greater than $1/2,$ hence physically acceptable, thus verifying the Heisenberg Uncertainty Principle.

\begin{figure}[H]
    \centering
    \includegraphics[width=\linewidth]{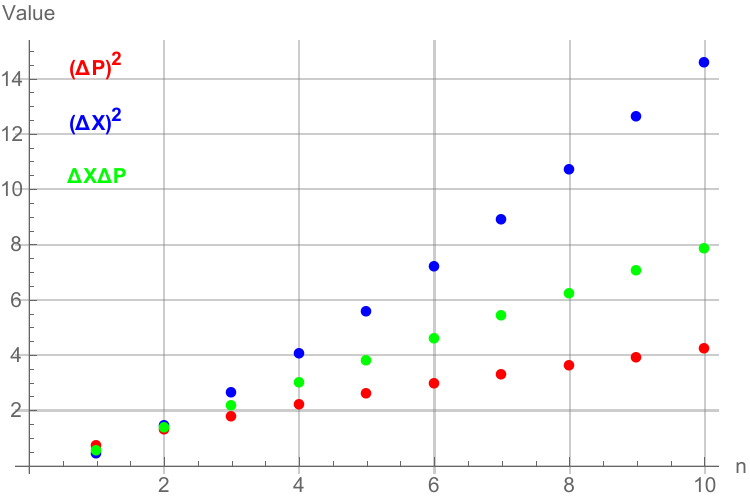} 
    \caption{ Graph Representing the Square of the Standard Deviation of Position ($X$), Momentum ($P$), and the Product of Uncertainties $\Delta X \Delta P$.}
    \label{fig:yourfigure}
\end{figure}

This graph illustrates how the uncertainties in position and momentum, as well as their product, vary with the index \(n\). This is consistent with the Heisenberg Uncertainty Principle, which states that the product of the uncertainties in position and momentum (\(\Delta X \Delta P\)) has a lower bound, typically \(\hbar/2\) (where \(\hbar\) is the reduced Planck constant).

\section{Conclusion}\label{sec13}

This study explores pseudo-Hermitian operators in quantum mechanics, expanding the boundaries of traditional Hermitian frameworks. By deriving the uncertainty relation for pseudo-Hermitian Hamiltonians and showing its equivalence to the standard Hermitian
uncertainty relation within the pseudo-inner product, we highlight the consistency and validity of this extended framework. 

As an application, we solved the time-dependent Schr\"{o}dinger equation for
the Hamiltonian of a particle with a time-dependent mass confined inside a
complex time-dependent triangular potential well. Using the invariant
method, we found exact solutions for this equation and calculated the
uncertainty relation, which was real and greater than 1/2. The uncertainty
relations for position and momentum operators confirm the effectiveness of
pseudo-Hermitian systems in accurately describing quantum phenomena.

\section*{Data Availability Statements}

The authors declare that the data supporting the findings of this study are
available in the article.

\end{document}